\documentstyle[12pt]{article}

\newcommand{\eq}[1]{(\ref{#1})}
\newcommand{\beq}{\begin{equation}}
\newcommand{\eeq}{\end{equation}}
\newcommand{\beqn}{\begin{eqnarray}}
\newcommand{\eeqn}{\end{eqnarray}}
\newcommand{\dd}{\mbox{d}}
\def\bbbone{{\mathchoice {\rm 1\mskip-4mu l} {\rm 1\mskip-4mu l}
{\rm 1\mskip-4.5mu l} {\rm 1\mskip-5mu l}}}
\def\bbbc{{\mathchoice {\setbox0=\hbox{$\displaystyle\rm C$}\hbox{\hbox
to0pt{\kern0.4\wd0\vrule height0.9\ht0\hss}\box0}}
{\setbox0=\hbox{$\textstyle\rm C$}\hbox{\hbox
to0pt{\kern0.4\wd0\vrule height0.9\ht0\hss}\box0}}
{\setbox0=\hbox{$\scriptstyle\rm C$}\hbox{\hbox
to0pt{\kern0.4\wd0\vrule height0.9\ht0\hss}\box0}}
{\setbox0=\hbox{$\scriptscriptstyle\rm C$}\hbox{\hbox
to0pt{\kern0.4\wd0\vrule height0.9\ht0\hss}\box0}}}}
\def\bbbe{{\mathchoice {\setbox0=\hbox{\smalletextfont e}\hbox{\raise
0.1\ht0\hbox to0pt{\kern0.4\wd0\vrule width0.3pt
height0.7\ht0\hss}\box0}}
{\setbox0=\hbox{\smalletextfont e}\hbox{\raise
0.1\ht0\hbox to0pt{\kern0.4\wd0\vrule width0.3pt
height0.7\ht0\hss}\box0}}
{\setbox0=\hbox{\smallescriptfont e}\hbox{\raise
0.1\ht0\hbox to0pt{\kern0.5\wd0\vrule width0.2pt
height0.7\ht0\hss}\box0}}
{\setbox0=\hbox{\smallescriptscriptfont e}\hbox{\raise
0.1\ht0\hbox to0pt{\kern0.4\wd0\vrule width0.2pt
height0.7\ht0\hss}\box0}}}}

\def\NP{ Nucl.~Phys.}

\def\PL{ Phys.~Lett.}
\def\PRL{ Phys.~Rev.~Lett.}
\def\cf{{\it cf.}}

\def\eg{{\it e.g.}}
\def\PRp{ Phys.~Rep.}
\def\PR{ Phys.~Rev.}
\def\etal{\hbox{\it et al.}}
\newcommand\Appendix[1]{\par
\setcounter{section}{0}
 \setcounter{equation}{0}
 \renewcommand{\thesection}{Appendix \Alph{section}}
\section{#1}
 \def\theequation{\Alph{section}.\arabic{equation}}}

\def\eg{{\it e.g.}}

\def\diff{\partial}

\def\cB{{\cal B}}
\def\cK{{\cal K}}
\def\cD{{\cal D}}
\def\dd{{\rm d}}
\def\half{\frac{1}{2}}

\title{
\vspace{-1.5cm}
\begin{flushright}
ITEP-94-63
\end{flushright}
\vspace{1.5cm}
Confinement Mechanism in Various Abelian Projections of $SU(2)$
Lattice Gluodynamics}

\author{M.N.~Chernodub
\thanks{ITEP and the Moscow Institute of Physics and Technology,
Dolgoprudny,\newline Moscow region, Russia}\\
M.I.~Polikarpov\\ and A.I.~Veselov\\{\sl ITEP, Moscow, 117259, Russia}}

\date{}

\begin{document}

\bibliographystyle{bibstand}

\maketitle
\begin{abstract}
We show that the monopole confinement mechanism in lattice gluodynamics is
a particular feature of the maximal abelian projection. We give an
explicit example of the $SU(2) \rightarrow U(1)$ projection (the minimal
abelian projection), in which the confinement is due to topological objects
other than monopoles. We perform analytical and numerical study of
the loop expansion of the Faddeev--Popov determinant for the maximal and the
minimal abelian projections, and discuss the fundamental modular region for
these projections.
\end{abstract}

\newpage

\section{Introduction}
In his well known--paper, 't Hooft \cite{tHo81} suggested a
partial gauge fixing procedure for the $SU(N)$ gluodynamics which
does not fix the $[ U(1)]^{N-1}$ gauge group. Under the abelian
transformations, the diagonal elements of the gluon field transform
as gauge fields; the nondiagonal elements transforms as matter fields. Due to
the compactness of the $U(1)$ gauge group, the monopoles exist, and if
they are condensed, the confinement of color can be explained in the
framework of the classical equations of motion \cite{Man76,tHo76}. The
string between the colored charges is formed as the dual analogue of the
Abrikosov string in a superconductor, the monopoles playing the role of
the Cooper pairs.

        Many numerical experiments (see \eg\ the review \cite{Suz93})
confirm the monopole confinement mechanism in the $U(1)$ theory obtained by
the abelian projection from the $SU(2)$ lattice gluodynamics.  The string
tension $\sigma_{U(1)}$ calculated from the $U(1)$ Wilson loops (loops
constructed only from the abelian gauge fields) coincides with the full
$SU(2)$ string tension \cite{SuYo90}; the monopole currents satisfy the
London equation for a superconductor \cite{SiBrHa93}. Recently it has been
shown \cite{ShSu94,StNeWe94} that the $SU(2)$ string tension is well
reproduced by the contribution of the abelian monopole currents.  Numerical
study of the effective monopole action \cite{ShSu294} shows that the entropy
of the monopole loops dominates over the energy, and therefore, there exists
the monopole condensate in the zero temperature $SU(2)$ lattice
gluodynamics. All these remarkable facts, however, have been obtained only
for the so called maximal abelian (MaA) projection
\cite{KrScWi87,KrLaScWi87}.  Other abelian projections (such as the
diagonalization of the plaquette matrix $U_{x,12}$) do not give evidence
that the vacuum behaves as the dual superconductor. Below we give two
relevant examples.

        First, it turns out \cite{IvPoPo90} that the fractal dimensionality
of the monopole currents extracted from the lattice vacuum by means of the
maximal abelian projection is strongly correlated with the string tension.
If monopoles are extracted by means of other projections, this correlation is
absent (\cf\ Fig.2 and Fig.4 of ref.\cite{IvPoPo90}). An other example is
the temperature dependence of the monopole condensate measured on the basis
of the percolation properties of the clusters of monopole currents
\cite{IvPoPo93}. For the maximal abelian projection the condensate is
nonzero below the critical temperature $T_c$ and vanishes above it. For the
projection which corresponds to the diagonalization of $U_{x,12}$, the
condensate is nonzero at $T>T_c$, and it is not the order parameter for the
phase transition. The last result has been obtained by the authors of
\cite{IvPoPo93}, but is unpublished.

        In the present publication we discuss the dependence of the
confinement mechanism on the type of the abelian projection.  We find that
the monopole confinement mechanism may be a particular property of the
MaA projection (section 2), and we give an explicit example of the abelian
projection \cite{PolChe94} in which confinement is due to topological
defects which are not monopoles (section 3). We also study the effective
$U(1)$ action in the considered projections, and show that it is rather
nonlocal (section 4).

\section{Maximal Abelian Projection and Compact Electrodynamics}

The MaA projection \cite{KrScWi87,KrLaScWi87} corresponds to the gauge
transformation that makes the link matrices diagonal ``as much as
possible''.  For the $SU(2)$ lattice gauge theory, the matrices of the gauge
transformation $\Omega_x$ are defined by the following the maximization
condition:

%\samepage{
\beq
	\max_{\{\Omega_x\}} R(U')\;, \label{MaAP}
\eeq

\beq
        R(U') = \sum_{x,\mu} Tr(U'_{x\mu}\sigma_3 U'^{+}_{x\mu}\sigma_3)\;,
      \  U'_{x\mu} = \Omega^+_x U_{x\mu} \Omega_{x+\hat{\mu}}\;. \label{RU}
\eeq
%}
For the standard parametrization of the $SU(2)$ link matrix, we have
$U^{11}_{x\mu} = \cos \phi_{x\mu} e^{i\theta_{x\mu}}; \ U^{12}_{x\mu} = \sin
\phi_{x\mu} e^{i\chi_{x\mu}}; \ U^{22}_{x\mu} = U^{11 *}_{x\mu}; \
U^{21}_{x\mu} = - U^{12 *}_{x\mu}; \ 0 \le \phi \le \pi/2, \ -\pi <
\theta,\chi \le \pi$; condition \eq{MaAP} has the form:

\beq
        \max_{\{\Omega_x\}}\sum_{x,\mu} \cos 2 \phi'_{x \mu}\;. \label{Mac}
\eeq
The $U(1)$ gauge transformations, which leave invariant the gauge conditions
\eq{MaAP}, \eq{Mac}, show that after the abelian projection $\theta$ becomes
the abelian gauge field and $\chi$ is the vector goldstone field, which
carry charge two in the continuum limit:

\beqn
        \theta_{x\mu} & \to & \theta_{x\mu} +\alpha_x -\alpha_{x+\hat{\mu}}
         \label{u1th}\;,\\
        \chi_{x\mu} & \to & \chi_{x\mu} +\alpha_x + \alpha_{x+\hat{\mu}}\;.
	  \label{u1chi}
\eeqn

It is instructive to consider the plaquette action in terms of the angles
$\phi, \ \theta $ and $\chi$:

\beq
S_P  =  \frac{1}{2}\mbox{Tr}\, U_1 U_2 U_3^+ U_4^+ = S^a + S^n + S^i\;,
\label{SP}
\eeq
where

\beqn
S^a  = & & \cos \theta_P\,
\cos\phi_1\, \cos\phi_2\, \cos\phi_3\, \cos\phi_4,
\nonumber \\
S^n  = & - & \cos (\theta_3 + \theta_4 - \chi_1 + \chi_2)\,
\cos\phi_3\, \cos\phi_4\, \sin\phi_1\, \sin\phi_2
\nonumber \\
 & + & \cos (\theta_2 + \theta_4 - \chi_1 + \chi_3)\,
                           \cos\phi_2\, \cos\phi_4\, \sin\phi_1\, \sin\phi_3
\nonumber \\
 & + & \cos (\theta_1 - \theta_4 + \chi_2 - \chi_3)\,
                           \cos\phi_1\, \cos\phi_4\, \sin\phi_2\, \sin\phi_3
\label{Sn} \\
 & + & \cos (\theta_2 - \theta_3 - \chi_1 + \chi_4)\,
                           \cos\phi_2\, \cos\phi_3\, \sin\phi_1\, \sin\phi_4
\nonumber \\
 & + & \cos (\theta_1 + \theta_3 + \chi_2 - \chi_4)\,
                           \cos\phi_1\, \cos\phi_3\, \sin\phi_2\, \sin\phi_4
\nonumber \\
 & - & \cos (\theta_1 + \theta_2 + \chi_3 - \chi_4)\,
                          \cos\phi_1\, \cos\phi_2\, \sin\phi_3\, \sin\phi_4,
\nonumber \\
 S^i  = & & \cos \chi_{\tilde{P}} \,
   \sin\phi_1\, \sin\phi_2\, \sin\phi_3\, \sin\phi_4; \nonumber
\eeqn
here we have set:

\beqn
\theta_P & =& \theta_1 + \theta_2 - \theta_3 - \theta_4\;, \label{P} \\
\chi_{\tilde{P}} & = & \chi_1 - \chi_2 + \chi_3 - \chi_4\;, \label{tP}
\eeqn
and the subscripts $1,...,4$ correspond to the links of the plaquette:  $1
\rightarrow \{x,x+\hat{\mu}\},...,4 \rightarrow \{x,x+\hat{\nu}$\}.  Note
that $S^a$ is proportional to the Wilson plaquette action of compact
electrodynamics for the ``gauge'' field $\theta$; the corresponding
action $S^i$ for the ``matter'' field $\chi$ contains the unusual
combination $\chi_{\tilde{P}}$ \eq{tP}, which is invariant under the gauge
transformations \eq{u1chi}. Action $S^n$ describes the interaction of the
fields $\theta$ and $\chi$.

         Due to condition \eq{Mac}, in the MaA projection the angle $\phi$
fluctuates about zero, and we can expect that the largest contribution to the
total action \eq{SP} comes from $S^a$, and  that $S^a > S^n > S^i$. This
conjecture is confirmed by numerical calculations. We use the standard heat
bath method to simulate $SU(2)$ gluodynamics on the
$10^4$ lattice, at 15 values of $\beta$; at each value of $\beta$ we used 15
field configurations separated by 100 of heat bath sweeps. To get obtain the
MaA projection, we performed 800 gauge fixing sweeps through the lattice for
each field configuration. It occurs that $<S^a>$ is close to the total
action, the maximal difference between $<S_P>$ and $<S^i>$ is at $\beta
\approx 2.2$, where $<S^a> \approx 0.82 <S_P>$; $S^i$ is unacceptably small:
$<S^i> \approx - 0.001 \pm 0.0004$ at $\beta = 2.2$, at other values of
$\beta$ the absolute value of $<S^i>$ is even smaller. It is clear that
if we neglect the fluctuations of the angle $\phi$, as well as the
Faddeev-Popov determinant, the $SU(2)$ action in the maximal abelian gauge
is well approximated by the $U(1)$ action: $S_P\approx\cos \theta_P$, with
the renormalized constant $\bar{\beta}=\beta  \cos^4 \phi$.

Since in
the compact electrodynamics the confinement is due to the monopole
condensation, it is not surprising that in numerical experiments the vacuum
of gluodynamics behaves in the MaA projection as the dual superconductor. Of
course, this is only an intuitive argument. The confinement in the $U(1)$
theory exists in the strong coupling region, in which the rotational
invariance is absent.  Therefore, in order to explain the confinement at
large values of $\beta$ in $SU(2)$ gluodynamics, we have to study in detail
some special features of the gauge fixing procedure (such as the
Faddeev-Popov--determinant, fluctuations of the angle $\phi$, etc.).
We discuss some of these questions in the section 4.

        The fact that $<S^a>$ is close to $<S_P>$ is very interesting; it
means that in the MaA projection there is a small parameter in the $SU(2)$
lattice gluodynamics, which is $\varepsilon = \frac{<S_P>-<S^a>}{<S_P>}$; at
all values of $\beta$, we have $\varepsilon \le 0.18$. The meaning of this
parameter is simple: it is the natural measure of closeness between the
diagonal matrices and the link matrices after the gauge projection.

\section{$SU(2)$ Gluodynamics in the Minimal Abelian Projection}

The minimal abelian (MiA) projection \cite{PolChe94} is defined similarly
to the MaA projection \eq{MaAP} by

\beq
        \min_{\{\Omega_x\}} R(U')\;, \label{MiAP}
\eeq
where $R(U')$ is defined by \eq{RU}. In this projection the largest part of
the plaquette action \eq{SP} is $S^i$, and the term which is most important
for the dynamics is $\cos \chi_{\tilde{P}}$ (rather than $\cos \theta_P$ as
it is in the MaA projection). The fields in the MiA projection can be
transformed into the fields in the MaA projection by the following
gauge transformation:

\beq
\Omega (x) = - i \sigma_2 \cdot \frac{(-1)^{x_1+x_2+x_3+x_4}+1}{2} +
\bbbone \cdot \frac{(-1)^{x_1+x_2+x_3+x_4} - 1}{2}\; . \label{Sigma2}
\eeq
Thus $\Omega(x)$ is equal to the unity in the
``odd'' sites of the lattice, and to $- i \sigma_2$ in the ``even'' sites;
this gauge transformation becomes singular in the continuum limit.
The angles
$\phi$, $\theta$ and $\chi$, which parametrize the link matrix $U_l$,
transform under this gauge transformation in the following way. If
the link starts at an even point, $\left((-1)^{x_1+x_2+x_3+x_4}=1\right)$,
then $U_l \rightarrow (-i \sigma_2) U_l$ and

\beq
\phi \rightarrow \frac{\pi}{2} - \phi, \ \theta \rightarrow -\chi, \ \chi
\rightarrow (\pi - \theta) \bmod 2\pi . \label{angl1}
\eeq
If the link starts at an odd point, then $U_l \rightarrow U_l (-i
\sigma_2)^+ $ and

\beq
\phi \rightarrow \frac{\pi}{2} - \phi, \ \theta \rightarrow (\pi + \chi)
\bmod 2\pi, \ \chi \rightarrow \theta . \label{angl2}
\eeq
Since $Tr(U'_{x\mu}\sigma_3 U'^{+}_{x\mu}\sigma_3) = \cos 2 \phi '$, it
follows that under this gauge transformation $Tr(U'_{x\mu}\sigma_3
U'^{+}_{x\mu}\sigma_3) \to - Tr(U'_{x\mu}\sigma_3 U'^{+}_{x\mu}\sigma_3)$,
and the fields in the MaA projection are transformed into fields in the MiA
projection (and vice versa). Moreover, the monopoles extracted from the
field $\theta$ in the MaA projection turn, in the MiA projection, into some
topological defects constructed from the ``matter'' fields $\chi$. We call
these topological defects ``minopoles''.

Minopoles can be extracted from a given configuration of gauge fields
similarly to monopoles: from the angles $\chi$ we construct gauge
invariant plaquette variables $\chi_{\tilde{P}} = \tilde{\dd}\chi \bmod 2\pi
$, where $\tilde{\dd}\chi$ is defined by \eq{tP}. From these plaquette
variables we construct the variables attached to the elementary cubes
${}^*j = \frac{1}{2\pi}\tilde{\dd} \chi_{\tilde{P}}$; for ${}^*j \neq 0$
the link dual to the cube carries the minopole current. We use the
notation $\tilde{\dd}$ (instead of $\dd$), since the gauge transformations
of $\chi$ given by \eq{u1chi} differ from the gauge transformations of
$\theta$ given by \eq{u1th}, and the construction of the plaquette variable
from the link variables and that of the cube variable from the plaquette
variables differ in an obvious way from the standard construction. For
example, $\dd\theta$ is defined by \eq{P} and $\tilde{\dd}\chi$ is defined
by \eq{tP}. In Fig.1 we illustrate the standard construction of the
monopoles from the field $\theta$, and the construction of the minopoles
from the field $\chi$.

Since monopoles, which exist in the MaA projection become minopoles in the
MiA projection, than if in the MaA projection the confinement phenomenon is
due to condensation of monopoles (constructed from the field $\theta$), then
in the MiA projection the confinement is due to other topological objects
(minopoles), constructed from the ``matter'' field $\chi$. We thus conclude
that {\em in the MiA projection the confinement is not due to monopoles and
the vacuum is not an analogue of the dual superconductor}. It should be
stressed that monopoles still exist in the MiA projection; they can be
extracted from the fields $\theta$ in the usual way, but they are not at all
related to the dynamics. To illustrate this simple fact we plot in Fig.2 the
space--time asymmetry of the {\bf \it monopole} currents
\cite{Bra91,Hio91}\footnote{The definition of this asymmetry
is obvious:
$A=<(J_t-J^S)/J_t>$, where $J^S= (J_x+J_y+J_z)/3$, $J_{\mu}$ is the monopole
current in the direction $\mu$.} for the $SU(2)$ gauge theory on the
$10^3\times4$ lattice for the MiA projection. In the same figure we also
show the asymmetry of the minopole currents in the MiA projection. These
results have been obtained by averaging over 10 statistically independent
field configurations for each value of $\beta$, and 500--800 of gauge
fixing sweeps have been performed for each configuration.

It is clearly seen that the asymmetry of the minopole currents is the order
parameter for the temperature phase transition, while the asymmetry of the
monopole currents is not. Since the monopole currents and the minopole
currents are interchanged when the fields are transformed from the MiA to
the MaA projection, Fig.2 also shows that  for the MaA projection the
asymmetry of the monopole currents is the order parameter, whereas the
asymmetry of minopole currents is not an order parameter.

        Minopoles are to some extend the lattice artifacts, since the gauge
fields in the MaA and the MiA projections are related by the gauge
transformation, which becomes singular in the continuum limit. We discuss
minopoles, since they clearly illustrate the dependence of the confinement
mechanism on the lattice, upon the type of the abelian projection.

\section{MaA and MiA Projections and the Lattice Path Integral}

There are a lot of facts (see Introduction) that in the MaA projection the
confinement is due to the monopole condensation. The confinement
mechanism is in some sense the same in the MiA projection, since the MiA
projection is related to the MaA projection by the global gauge
transformation \eq{Sigma2}. Therefore it is important to study the effective
$U(1)$ action in the MaA (and in the MiA) projection and in this section we
discuss the MiA and the MaA projections in the lattice path integral. It is
easy to find that the Faddeev--Popov determinant is {\bf \it the same} for
the MaA and the MiA projections, the difference between them being in the
region of integration over the gauge fields; this region is called the
``Fundamental Modular Region'' (FMR) \cite{FMR}.  In Appendix we briefly
describe how the FMR appears in the gauge fixing procedure. The gauge fixing
actions $S_{gf}$ for the MaA and the MiA projections \eq{MaAP}, \eq{MiAP}
are given by:

\beq
        S^{a}_{gf}(U) = 2 - R(U),
        S^{i}_{gf}(U) = 2 + R(U)\;, \label{GFAct}
\eeq
and $R(U)$ is defined by \eq{RU}.
It is clear that $S^a_{gf}$ and $S^i_{gf}$ transform into each other by the
local gauge transformation \eq{Sigma2}. From the definition of
the FMR, given by eq.\eq{DefFMR}, we conclude that all fields from the FMR
for the MaA projection transform, by this transformation, into fields which
lie in the FMR for the MiA projection and vice versa. Moreover, the
determinants of the matrices $\cD$ \eq{ExpAct} coincide for both
projections; and the final expression for the partition function for the
fixed projection \eq{FixedPartFunct2} differs by the FMR. It is remarkable
that the difference in the integration regions leads to different
confinement mechanisms in these projections.

There exist a natural loop expansion of the Faddeev--Popov determinant;
below we calculate two leading terms of this expansion. For definiteness,
consider the MaA projection. Expanding the MaA gauge fixing action
\eq{GFAct} in powers of $\omega$ (see eq.\eq{ExpAct}) we obtain the
stationary point condition \eq{C eq 0}:

\beq
 {\cal C}^a_x(W) = \sum_{b=1,2} \sum^4_{\mu=1}
 {\bigl[\tilde{U}_{x,\mu}\sigma^3 \tilde{U}^{+}_{x,\mu} +
 \tilde{U}^{+}_{x,-\mu}\sigma^3 \tilde{ U}_{x,-\mu} \bigr]}^b
 \epsilon^{ab3} = 0\; . \label{FixEqs}
\eeq

The corresponding gauge fixing equations in the continuum limit are
\cite{KrScWi87,KrLaScWi87}:
$\sum_{\mu} (A^3_{\mu} \pm i \diff_{\mu}) A^{\pm}_{\mu} = 0$,
$A^{\pm}_{\mu} = A^1_{\mu} \pm i A^2_{\mu}$.

      The Faddeev--Popov operator $\cD$, which enters the expansion
\eq{ExpAct}, is given by (up to irrelevant constant):

\beq \cD^{ab}_{x,y} = 2 D \cK^{ab}_{x,y}(U) + \cB^{ab}_{x,y}(U)\;,
        \label{FPOperator}
\eeq
where
\beq
        \cK^{ab}_{x,y}(U) =
  \tilde\cK_x\,\delta_{xy}\,(\delta^{ab} - \delta^{a3}\delta^{b3}), \quad
  \tilde\cK_x = \frac{1}{2 D} \sum_{\mu= - D}^{D}
  \cos(2\phi_{x,x+\hat{\mu}});
\label{Kxy}
\eeq
here $D$ is the dimension of the space--time;
the angle $\phi_{x,y}$ is one of the parameters, defining the link
matrix (see eq.\eq{Mac}); and
\beq
        \cB^{ab}_{x,x+\hat{\mu}}(U) =
        Tr\Bigl[\sigma^a U_{x,x+\hat{\mu}} \sigma^b
        U^+_{x,x+\hat{\mu}}\Bigr] \,\epsilon^{ac3}\;.
        \epsilon^{bd3}
\eeq
The indices $x,\ y$  correspond to sites of the lattice and $a, \ b$
are the color indices.
The matrix $\cK^{ab}_{x,y}$ is diagonal with respect to the spatial
indices $x$, $y$; the sum in \eq{Kxy} is taken over all links
connected to the site $x$.  The matrix $\cB^{ab}_{xy}$ is nonzero if the
points $x$ and $y$ belong to the beginning point and to the end--point of a
link.

The determinant of the matrix $\cD$ can be represented as the exponential
function of the sum over all closed graphs $\{L\}$ of length $L$:

\beqn
        {Det}^{\half}\Bigl\{\cD(U)\Bigr\} =
\nonumber \\
        const.\exp\Bigl\{ \sum_x \ln(|\tilde\cK_x|) -
        \sum^{\infty}_{L=2} \frac{1}{2 L} {\Bigl( -
        \frac{1}{2 D}\Bigr)}^{L} \sum_{l\in\{L\}} Tr
        \prod_{l}{\cK}^{-1}\cB
        \Bigr\}, \label{ExplicitFP}
\eeqn
where the product $\prod_{l}$ is along the path $l$.
Since any closed path consists of the even number of links, this expansion is
in powers $\frac{1}{D^2}$. We define the effective action as:
$e^{-S}{Det}^{\half}\{\cD(U)\}=e^{-S_{eff}}$. The loop expansion of the
effective action is $S_{eff}=\beta \sum_P S_P + S^{(0)} + S^{(1)} + S^{(2)}
+ O(\frac{1}{D^6})$. Here $S^{(k)}$ corresponds to the loop of  length $2k$,
and has the order $O(\frac{1}{D^{2k}})$:

\beq
        S^{(0)} = - \sum_x \ln(|\tilde\cK_x|),  \label{S0}
\eeq
$S^{(1)}$ corresponds to the loop, belonging to one link:

\beq
       S^{(1)} = \sum_{x,\mu} \frac{3 +  \cos 4 \phi_{x\mu}}{4 D^2
       \tilde\cK_x \tilde\cK_{x+\mu}}\,. \label{S1}
\eeq

It is natural to subdivide the loops which contribute to $S^{(2)}$ into two
types ($S^{(2)}=S^{(2,1)} + S^{(2,2)}$): the one--dimensional loops passing
through the points $x$, $x+\mu$, $x$, $x+\nu$ and returning
to the point $x$, and the loops which correspond to plaquettes. The
contribution of the one--dimensional loops is:

\beqn
       S^{(2,1)} = \sum_{x} \sum^D_{\mu,nu=-D}
       \frac{1}{64 D^4 \tilde\cK^2_x \tilde\cK_{x+\mu} \tilde\cK_{x+\nu}}
       \Bigl[(3 + \cos4\phi_{x\mu})
       (3 + \cos4\phi_{x\nu}) + \nonumber\\
       4 \sin^2 2\phi_{x\mu} \sin^2 2\phi_{x\nu}
       \cos(2(\chi_{x\mu}-\chi_{x\nu}+\theta_{x\mu}-\theta_{x\nu}))\Bigr]\,.
       \label{S21}
\eeqn

Note that in this expression we have $\mu = 1,...,4$ and
$\nu=-4,...,-1,1,...,4$.  If $\nu=\mu$, then loop belongs to a single link;
if $\nu =- \mu$, then the loop corresponds to the straight line; and if
$|\nu| \neq \mu$, then the loop corresponds to two neighboring links,
perpendicular to each other.

The effective potential of the field $\phi$, which corresponds to
$S^{(0)} + S^{(1)}$,\ has its minimum at the points $\phi = 0$ ($\phi =
\pi\slash 2$), and it tends to infinity as $\phi$ approaches $\pi\slash4$.
Thus the field $\phi$ fluctuates about the value $\phi = 0$ ($\phi =
\pi\slash 2$) for the MaA (MiA) projection.

The term corresponding to the plaquette loop in the expansion
\eq{ExplicitFP} defines the correction to the plaquette action \eq{SP}:

\beq
  S^{(2,2)} = \delta S^{a} + \delta S^{n} + \delta S^{i}\;, \label{S22}
\eeq
where

\beqn
\delta S^{a}=
&&C_P(\phi) \cos 2 \theta_P\, \cos^2\phi_1\,
\cos^2\phi_2\, \cos^2\phi_3\, \cos^2\phi_4,
\nonumber \\
\delta S^{n} = && \nonumber \\
 && C_P(\phi) [ \cos(2 (\theta_3 + \theta_4 - \chi_1 + \chi_2))\,
\cos^2\phi_3\, \cos^2\phi_4\, \sin^2\phi_1\, \sin^2\phi_2
\nonumber \\
 & + &  \cos(2 (\theta_2 + \theta_4 - \chi_1 + \chi_3))\,
        \cos^2\phi_2\, \cos^2\phi_4\, \sin^2\phi_1\, \sin^2\phi_3
\nonumber \\
 & + & \cos(2 (\theta_1 - \theta_4 + \chi_2 - \chi_3))\,
        \cos^2\phi_1\, \cos^2\phi_4\, \sin^2\phi_2\, \sin^2\phi_3
\label{Sn1} \\
 & + & \cos (2 (\theta_2 - \theta_3 - \chi_1 + \chi_4))\,
        \cos^2\phi_2\, \cos^2\phi_3\, \sin^2\phi_1\, \sin^2\phi_4
\nonumber \\
 & + & \cos (2 (\theta_1 + \theta_3 + \chi_2 - \chi_4))\,
        \cos^2\phi_1\, \cos^2\phi_3\, \sin^2\phi_2\, \sin^2\phi_4
\nonumber \\
 & + & \cos (2 (\theta_1 + \theta_2 + \chi_3 - \chi_4))\,
        \cos^2\phi_1\, \cos^2\phi_2\, \sin^2\phi_3\, \sin^2\phi_4] ,
\nonumber \\
\delta S^{i}  = & & C_P(\phi) \cos (2 \chi_{\tilde{P}})\,
     \sin^2\phi_1\, \sin^2\phi_2\, \sin^2\phi_3\, \sin^2\phi_4; \nonumber
\eeqn
$\theta_P$, $\chi_{\tilde{P}}$ and $\tilde\cK_x(\phi)$ are given by eqs.
\eq{P}, \eq{tP} and \eq{Kxy};

\beq
 C_P(\phi) = \frac{2}{ D^4}\, \frac{1}{\tilde\cK_{x_1}(\phi)\,
 \tilde\cK_{x_2}(\phi)\, \tilde\cK_{x_4}(\phi)\,
 \tilde\cK_{x_4}(\phi)}.
\eeq
As in \eq{Sn}, the subscripts $1,...,4$ of the angles $\phi$, $\theta$,
$\chi$ correspond to four links of the plaquette under consideration, and
the points $x_1$...$x_4$ are the corners of this plaquette. Note that
$S_{eff}$ is invariant under the $U(1)$ gauge transformations \eq{u1th},
\eq{u1chi}, as it should be. In addition to the $U(1)$ action for the
fundamental representation $S^a \sim \cos \theta_P$ the adjoint action
$\delta^{(1)} S^{a} \sim\cos 2 \theta_P$ exists in the effective action.

In order to study the effective $U(1)$ action,
we calculated
numerically the quantum averages $<S^{(0)}>$, $<S^{(1)}>$ and $<S^{(2)}>$
in the MaA projection. The dependence of these quantities on $\beta$ is shown
in Fig.3. These results have been obtained by averaging over 10
statistically independent field configurations for each value of $\beta$,
500--800 of the gauge fixing sweeps have been performed for each
configuration. The asymptotics of $S^{(i)}$, shown in this figure, are
given by the formulae:

\beq
        <S^{(0)}> = \frac{1}{12\beta} + O(\frac{1}{\beta^2})
        \nonumber
\eeq

\beq
        <S^{(1)}> = \frac{1}{24}\left(1+\frac{1}{2 \beta}\right)
         + O(\frac{1}{\beta^2}) \label{Sasym}
\eeq
\beq
        <S^{(2)}> = \frac{1}{96}\left(\frac{7}{4}+\frac{1}{\beta}\right)
         + O(\frac{1}{\beta^2}) \nonumber
\eeq

These expressions can be easily found from \eq{S0}--\eq{Sn1} by the
standard low temperature expansion technique. For their derivation we have
used the fact that $\frac{<S^{n}>}{<S^{a}>} = O(\frac{1}{\beta^2})$ (we can
not prove this fact analytically but our numerical data clearly confirm it).

{}From Fig.3 we conclude that the higher loops in the effective action are not
strongly suppressed; the same conclusion was maid in \cite{Suz93},
\cite{Yee94}. In these papers the effective $U(1)$ action integrated over
$\phi$ and $\chi$ was studied numerically.

\section{Conclusions and Acknowledgments}

        If monopoles are responsible for the confinement in MaA projection
and minopoles are responsible for the confinement in the MiA projection,
what are the important topological excitations in a general abelian
projection? If both diagonal and nondiagonal gluons are not suppressed,
then string--like topological defects can also be important for the dynamics
of the system \cite{ChPoZu94}. The idea is: nondiagonal gluons transform
under the $U(1)$ gauge transformations as matter fields, diagonal gluons
transform as gauge fields, and an analogue of the Abrikosov--Nielsen--Olesen
strings exists in gluodynamics after the abelian projection. Between
strings made of condensed nondiagonal gluons (which carry the $U(1)$ charge
{\bf 2}) and the test quark of the charge {\bf 1}, there exists topological
interaction \cite{AhBheff,PoWiZu93}, which is the analogue of the Aharonov
-- Bohm effect. Thus, in the effective $U(1)$ action of the $SU(2)$
gluodynamics there probably exists a very specific topological interaction.
We describe an analytical and numerical study of this interaction in a
separate publication.

The
topological defects discussed above may be a reflection of some $SU(2)$ gauge
field configuration. For example, monopoles and minopoles may be the abelian
projection of $SU(2)$ monopoles \cite{SmvdSi93}. In ref. \cite{TrPoWo93} it
is found that the ``extended monopoles'' \cite{IvPoPo90} may be important
for the confinement mechanism in different abelian projections of the $3D$
$SU(2)$ gluodynamics. Finally, we note that it was found recently
\cite{SuIlMaOkYo94} that the contribution of the Dirac sheets to the
abelian Polyakov loops plays the role of the order parameter for finite
temperature lattice gluodynamics; it is interesting that this fact holds
not only for the MaA projection but also for others unitary gauges.
Note that in the MiA projection monopoles are substituted by minopoles, and
in this projection we expect that the contribution of the minopole ``Dirac
sheets'' is correlated with the expectation values of the Polyakov loops.

MIP is grateful to E.~Akhmedov, P.~van~Baal, R.~Haymaker, J.~Smit and K.~Yee
for interesting discussions. This work was supported by the grant number
MJM000, financed by the International Science Foundation, by the JSPS
Program on Japan -- FSU scientists collaboration and by the grant number
93-02-03609, financed by the Russian Foundation for the Fundamental Sciences.

\Appendix{The Gauge Fixing Procedure
And The Fundamental Modular Region}

Here we briefly describe how the Fundamental Modular Region \cite{FMR}
arises in the
gauge fixing procedure. The standard method of gauge fixing is to
substitute the ``Faddeev -- Popov unity''

\beq
        1 = \Delta_{FP}(U) \int [\dd \Omega] \exp\Bigl\{- \lambda S_{gf}
(\Omega U \Omega^{+})\Bigr\}                \label{FPUnity}
\eeq
into the path integral

\beq
        {\cal Z} = \int [\dd U] \exp\bigl\{- S(U)\bigr\},   \label{PathInt}
\eeq
the limit $\lambda \rightarrow + \infty$ is implied.
The gauge fixing actions $S_{gf}$ for the MaA and the MiA projections are
given by \eq{GFAct}. Since $\lambda \rightarrow + \infty$ in \eq{FPUnity},
we can use the saddle point approximation to calculate $\Delta_{FP}(U)$. To
this end, we parametrize $\Omega$ in \eq{FPUnity} in the following way:
$\Omega(\omega) = {\Omega}^0_x \exp[\frac{i\sigma^a}{2} \omega^a_x]$,
where $\Omega^0_x$ is such that the fields $\tilde{U} = \Omega^0_x
U_{x,x+\hat{\mu}} \Omega^{0 +}_{x,x+\hat{\mu}}$ correspond to the absolute
minimum of $S_{gf}$, so that $\Omega^0 = \Omega^0(U)$. At the saddle point
we can restrict ourselves to small values of $\omega = \omega^a_x$, and

\beq
        S_{gf}\bigl(\Omega(\omega) U \Omega^+(\omega)\bigr) =
        S_{gf}(\tilde{U}) + {\cal C}\omega + \half \omega
        \cD \omega + O(\omega^3), \label{ExpAct}
\eeq
where ${\cal C} = {\cal C}^{a}_{x}(\tilde{U})$ and
$\cD = \cD^{ab}_{xy}(\tilde{U})$. Since $\tilde{U}$
corresponds to the absolute minimum of $S_{gf}$, we have

\beq
{\cal C}^{a}_{x}(\tilde{U}) = 0. \label{C eq 0}
\eeq
Substituting \eq{ExpAct} into \eq{FPUnity} and integrating over $\omega$, we
get the standard expression for the Faddeev -- Popov determinant:
$\Delta_{FP}(\tilde{U}) = const. \exp\Bigl\{ \lambda S_{gf}(\tilde{U})
\Bigr\}\, {Det}^{\half}\,\Bigl\{\cD(\tilde{U})\Bigr\}$. Substituting
\eq{FPUnity} into the path integral \eq{PathInt} and using the gauge
invariance of $S(U)$, we get the product of the group volume $\int [\dd
\Omega]$ and the partition function for the fixed gauge:

\beq
        {\cal Z} = \int [\dd U] \exp\Bigl\{- S(U) + \lambda\Bigl[S_{gf}
        (\tilde{U}) - S_{gf}(U)\Bigr]\Bigr\}
        Det^{\half}\Bigl\{\cD(\tilde{U})\Bigr\}\;.
        \label{FixedPartFunct2}
\eeq
Using once again the fact that $\lambda \rightarrow + \infty$,
we see that the nonzero contribution to this path integral is given by
$U$ belonging to the global minima of $S_{gf}(U)$:

\beq
        \Lambda = \Bigl\{U\ :\ S_{gf}(U) \leq
    S_{gf}(\Omega U \Omega^+)\ \forall\ \Omega \Bigr\}\;.   \label{DefFMR}
\eeq
By  definition,   in  this   region  we   have  $\Omega^0(U)   =  1   $   and
$S_{gf}(\tilde{U}) \equiv S_{gf}(U)$. The global minimum of the gauge  fixing
action  is  usually  called  the  $Fundamental  \  Modular  \  Region$  (FMR)
\cite{FMR}. In  order to  restrict the  integration over  the gauge fields to
this region, the step function

\beq
        \Gamma_{FMR}(U) = \cases{1&if $U\in\ \Lambda$;\cr 0&otherwise}\;,
\eeq
should be inserted \cite{FMR}; into the path integral
\eq{FixedPartFunct2} and
we thus obtain the following form of the partition function for the fixed
gauge:

\beq
        {\cal Z} = \int [\dd U] \exp\{- S(U)\} Det^{\half} \Bigl\{
        \cD(U) \Bigr\} \Gamma_{FMR}(U)\;.
                \label{FixedPartFunct3}
\eeq
To get this expression we have used the fact that $\Omega^0(U) = 1$ if $U$
lies in the FMR.

%\bibliography{mypapers,lattice}
%\end{document}

\newpage
\vspace{1cm}
\section*{Figure captions}
Fig.1. Construction of a monopole from the field $\theta$ and minopole from
the field $\chi$.\\
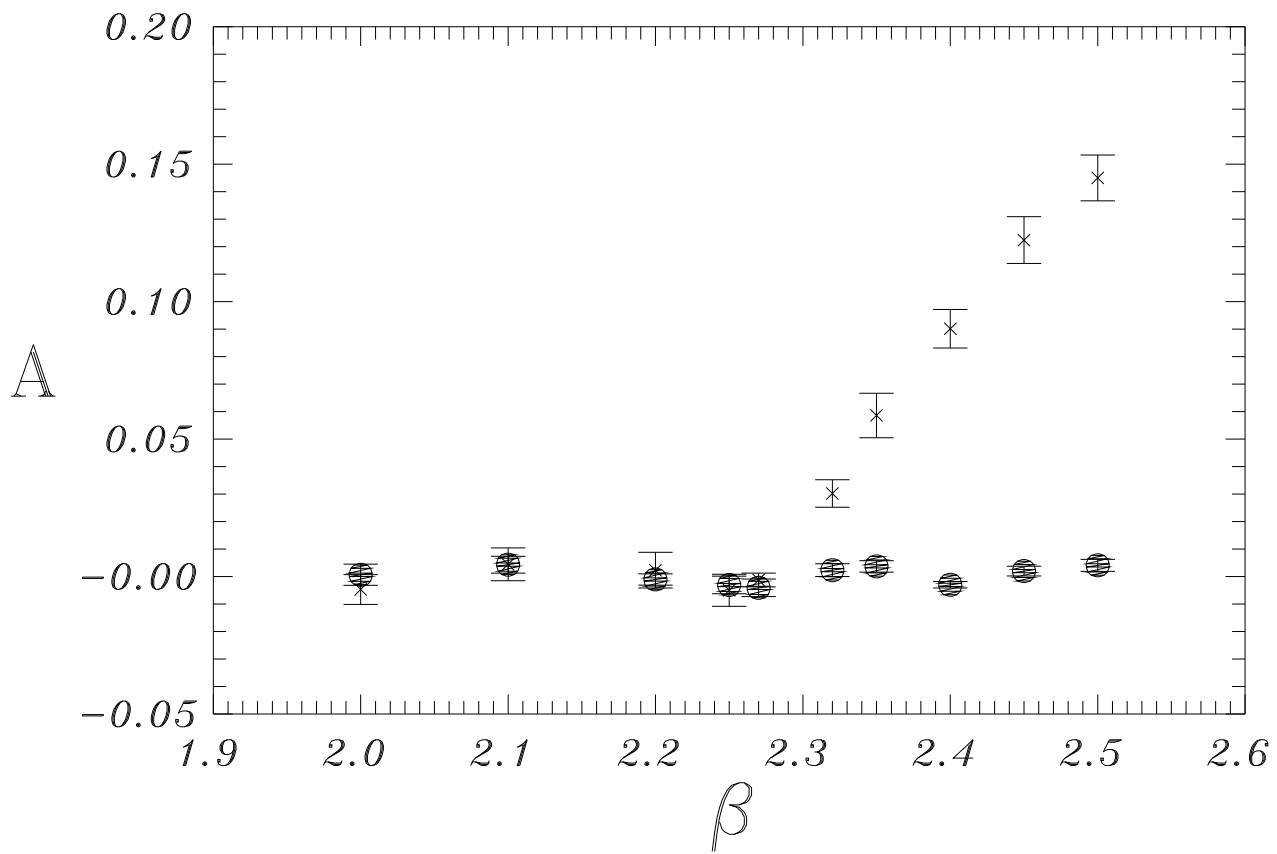
Fig.2. Asymmetry of the monopole currents (circles) and the minopole
currents (crosses) in the minimal abelian projection.\\
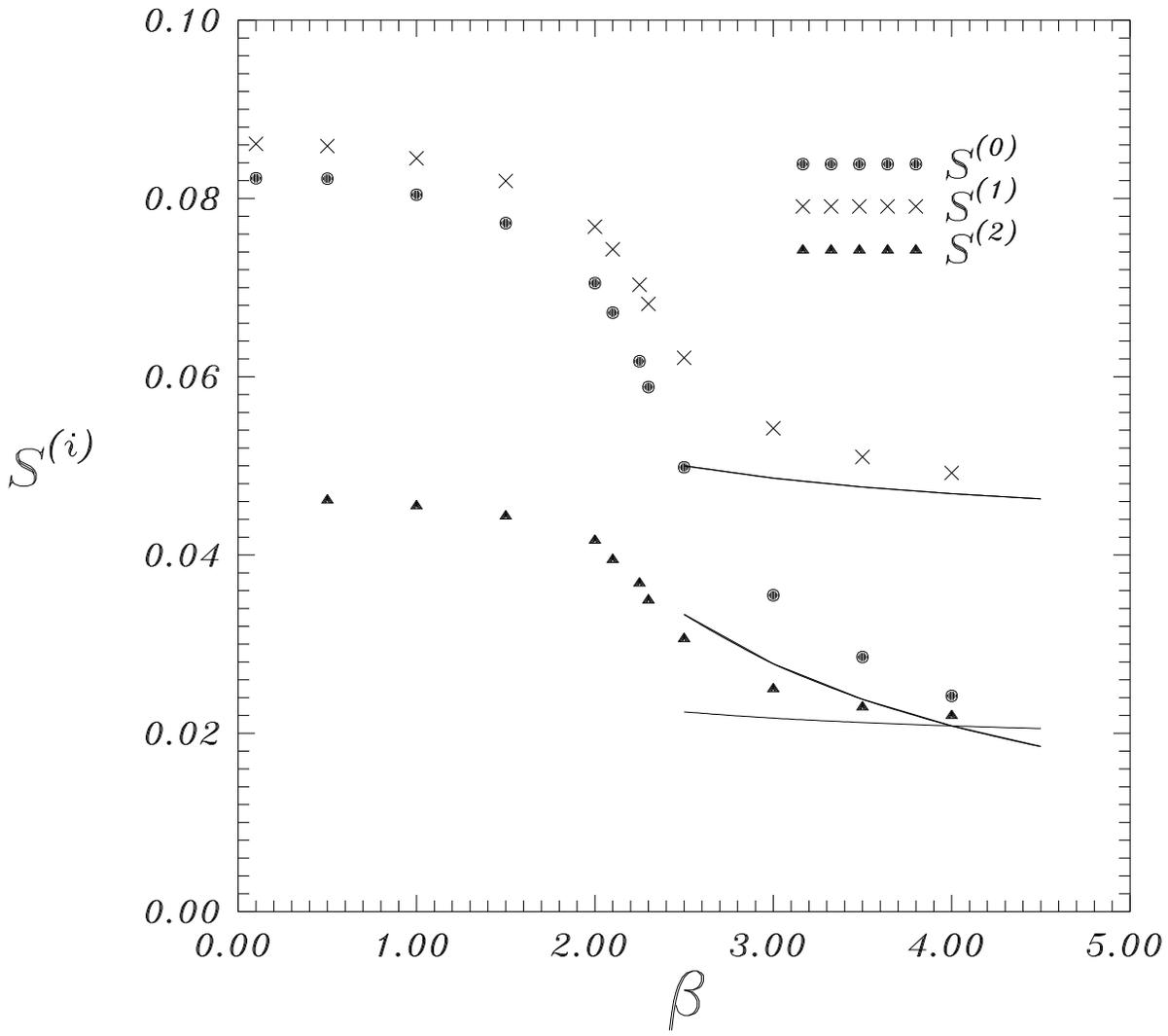
Fig.3. The corrections to the plaquette action, $S^{(i)}$, versus $\beta$.
The asymptotics shown in this figure, are
given by the formulae \eq{Sasym}.
Estimated statistical errors are of the size of the graphical symbols.

\end{document}